\documentclass{aa}
\usepackage{times}
\usepackage{graphics}
\usepackage[dvips]{epsfig}
\begin{document}
\thesaurus{04(12.07.1,12.04.1,08.12.2,10.08.1,08.16.2)}
\title{Estimating physical quantities for an observed galactic microlensing event}
\author{M. Dominik}
\institute{Institut f\"ur Physik, Universit\"at Dortmund, D-44221 Dortmund, Germany}
\date{Received ; accepted}
\maketitle
\begin{abstract}
For a given spatial distribution of the lenses and 
distribution of the transverse velocity of the lens
relative to the line-of-sight, 
a probability distribution for the lens mass for a single 
observed event is derived. 
In addition, similar
probability distributions are derived for the Einstein radius and the
separation of the lens objects
and their rotation period for a binary lens. 
These probability 
distributions are distinct from the distributions for the lens population,
as investigated e.g. by the mass moment method of De R{\'u}jula et al.\ 
(\cite{RJM}). It is shown that the expectation
value for the mass of a certain event as derived in this paper coincides 
with the estimated average mass of the underlying mass spectrum as found with 
the mass moment method when only one event is considered. 
The special cases of a Maxwellian velocity distribution and
of a constant velocity  are discussed in detail. For a rudimentary model of 
the Galactic halo, the probability distributions are shown and the relations
between the expectation values of the physical quantities and the event 
timescale are given. For this model it is shown that within a 
95.4~\%-interval around the expectation value the mass 
varies by a factor of 800. For the observed events towards the LMC --- including the binary lens models for MACHO LMC\#1 (Dominik \& Hirshfeld \cite{dohi2}) 
and MACHO LMC\#9 (Bennett et al.\ \cite{bennett1}) --- the results are shown explicitly.
I discuss what information can be extracted and how additional information
from the ongoing microlensing observations influences the results.

\keywords{gravitational lensing --- dark matter --- Stars: low-mass, brown
dwarfs --- Galaxy: halo --- planetary systems}
\end{abstract}

\section{Introduction}
Some attempts have been made to obtain information about the mass of the 
lens from observed microlensing events. Unfortunately, the mass cannot be
inferred directly. Instead, the only relevant 
information directly available from
a fit of a light curve to the observed data points is the timescale
$t_\mathrm{E}$, and the mass depends on this timescale ($M \propto t_\mathrm{E}^2$)
as well as on the
position of the lens and the transverse velocity of the lens
relative to the line-of-sight. Since the latter parameters are
both not observable (except for extraordinary cases where
some additional information can be obtained), one can only obtain
statistical information on the lens mass assuming distributions of the lens
position and the transverse lens velocity. Griest (\cite{griest}),
cited as GRI in the following,
argued that the most likely mass distribution is that which yields the
best fit to the distribution of timescales of the observed events. 
De R{\'u}jula et al.\ (\cite{RJM}), cited as RJM in the following,
Jetzer \& Mass{\'o} (\cite{JM}) and Jetzer (\cite{jetzer}) have shown that one can extract 
statistical moments of the lens mass distribution from the moments of the
distribution of the timescales. 

While these attempts use the distribution of
timescales to obtain information about the mass spectrum of the lenses
as realized in nature, I will discuss
another topic in this paper. For a given observed event, I investigate the
probability distributions of the lens mass and other physical quantities like
the Einstein radius and the separation and rotation period 
of the lens objects for a binary lens. These probability distributions 
give the
answer to the question, how probable certain ranges of the considered
physical quantity are for the given lens system 
having produced the event. Note that this question
is not answered by applying the methods of GRI or RJM. It is however
of special importance for planetary
systems. Also note that additional parameters will enter the calculation of
the event rate if one considers events which deviate from the 
point-source-point-mass-lens model. A binary lens will give serious
problems for determining the mass spectrum from the timescale distribution
since one measures the agglomeration of two objects each from the mass
spectrum for one event rather than a single object from the mass spectrum.
In contrast to this, the probability distributions presented in this paper
also give meaningful results for `anomalous' events.

This paper is organized as follows. In Sect.~2, it is shown which information
directly results from a fit of a galactic microlensing light curve.
In Sect.~3, the relation between the event rate and the mass spectrum,
the spatial distribution of the lenses, and the distribution of 
the relative velocity is given. Section~4 shows how the probability distribution
for the lens mass can be derived. Section~5 gives results for the moments
of further physical quantities and the probability distribution 
of these quantities around their expectation value. In Sect.~6, it is shown
that the expectation value for the mass coincides with the value obtained
by applying the mass moment method of RJM to one event. 
In Sect.~7, two special forms of the velocity distribution are discussed 
in more detail: a Maxwellian distribution and a fixed velocity. In Sect.~8,
the expectation values and the probability distributions for the lens mass,
the Einstein radius, and the separation and the rotation period for binary 
lenses are shown for a simple model of the galactic halo. In addition,
intervals corresponding to probabilites of 68.3~\% and 95.4~\% are given
for these quantities.
In Sect.~9, the implications for the observed events towards the LMC
(Alcock et al.\ \cite{alcock1}; Auborg et al.\ \cite{aubourg}; Dominik \& Hirshfeld \cite{dohi2};
Alcock et al.\ \cite{alcock2}, \cite{alcock4}; Pratt et al.\ \cite{pratt}; 
Bennett et al.\ \cite{bennett1}) are discussed explicitly. 
I discuss what information can
be extracted and how the observation of more events influences the results.

\section{Information from a fit of a galactic microlensing event}
\label{chap:estpar}
From a fit of the light curve to the observed data of a galactic microlensing
event, the only dimensional parameters are the point of time
$t_\mathrm{max}$ when the event occurs and the characteristic timescale 
$t_\mathrm{E}$ related to its duration. While the point of time $t_\mathrm{max}$
does not yield any relevant information, all physical quantities related to
the observed event which involve a dimension depend on $t_\mathrm{E}$. This is
true not only for the `standard model' of Galactic microlensing ---
a point-mass lens and a point source ---, but also for `anomalous' events
(binary sources and lenses, parallax effects, blending, finite size of the
source and the lens).\footnote{Since these `anomalous' events correspond to
a more general model and the point-source-point-mass-lens model is a special
case or an approximation, the `anomalous' events are quite the normal thing!}
The geometry of the microlensing events depends on a length scale which can
be chosen as the Einstein radius $r_\mathrm{E}$ of the lens of mass $M$ at a 
distance $D_\mathrm{d}$ from the observer, where the source is at a distance
$D_\mathrm{s}$ from the observer and at a distance $D_\mathrm{ds}$ from the lens.
The Einstein radius $r_\mathrm{E}$ is then given by
\begin{equation} 
r_\mathrm{E} = \sqrt{\frac{4GM}{c^2}\,\frac{D_\mathrm{d} D_\mathrm{ds}}{D_\mathrm{s}}}\,.
\end{equation} 
With $\mu = M/M_{\sun}$ and $x = D_\mathrm{d}/D_\mathrm{s}$, $r_\mathrm{E}$ can be
written as
\begin{equation}
r_\mathrm{E} = r_0\,\sqrt{\mu x(1-x)}\,
\end{equation}
where
\begin{equation}
r_0 = \sqrt{\frac{4GM_{\sun} D_\mathrm{s}}{c^2}}\,.
\end{equation}
The characteristic time scale $t_\mathrm{E}$ is given by
\begin{equation}
t_\mathrm{E} = \frac{r_\mathrm{E}}{v_{\perp}}\,,
\end{equation}
where $v_{\perp}$ is the transverse velocity of the lens relative to the
line-of-sight source-observer. Note that motions of the source and the
observer are also absorbed into this quantity.

For a given $t_\mathrm{E}$, the mass $\mu$ therefore follows as
\begin{equation}
\mu = \frac{t_\mathrm{E}^2}{r_0^2}\,\frac{v_{\perp}^2}{x(1-x)}
= \frac{t_\mathrm{E}^2 v_\mathrm{c}^2}{r_0^2}\,\frac{\zeta^2}{x(1-x)}\,,
\label{addeq}
\end{equation}
where $v_\mathrm{c}$ is a characteristic velocity and $\zeta = 
v_{\perp}/v_\mathrm{c}$.
One sees that $\mu$ depends on the timescale $t_\mathrm{E}$ as well as on
$x$ and $\zeta$. By assuming distributions of $x$ and $\zeta$ (where the
distribution of $\zeta$ may depend on $x$), it should in principle be
possible to derive a probability distribution for $\mu$. For doing this,
I have a look at the event rate in the next section.

\section{Event rate and mass spectrum}
\label{estpar:evrate}
Consider a coordinate system where the lens is at rest and let the
source move on a straight line projected onto the lens plane with
velocity $v_{\perp}$. Following Mao \& Paczy{\'n}ski (\cite{MP1}),
the characteristic width $w$ is then defined as the range of impact 
parameters for which a microlensing event occurs. Clearly, the width
$w$ is proportional to the Einstein radius $r_\mathrm{E}$, so that 
$w = w_0 r_\mathrm{E}(x)$. The event rate $\Gamma$ is 
given by the product of the area number density of the 
lenses, the
perpendicular velocity and the characteristic width of the considered 
type of event:
\begin{equation}
\Gamma = n\,v_{\perp}\,w\,.
\end{equation}

For variable lens position, the area number density 
of the lenses has to be replaced by an
integral of the volume number density $n$ 
over the line-of-sight direction $x$. 
For a general lens population, the number density depends also on the 
mass $\mu$ of the considered objects, so that one gets
\begin{equation}
D_\mathrm{s} \int \frac{dn(x,\mu)}{d\mu}\,d\mu\,dx
\end{equation}
as area number density of the lenses. 
If the mass spectrum does not depend on $x$, one
can separate the $x$ and $\mu$-dependence by
\begin{equation}
\frac{dn(x,\mu)}{d\mu} = H(x)\,\frac{dn_0(\mu)}{d\mu}\,,
\end{equation}
where the function $H(x)$ follows the volume mass density $\rho(x)$ as
$\rho(x) = \rho_0 H(x)$, so that $\rho(x) = \rho_0$ at the reference distance
where $H(x) = 1$.

The total volume number density of lenses at the reference distance is 
\begin{equation}
n_0 = \int \frac{dn_0(\mu)}{d\mu}\,d\mu\,,
\end{equation}
so that the probability for a mass $\mu$ in the interval $[\mu,\mu+d\mu]$ is
\begin{equation}
\omega(\mu)\,d\mu = \frac{1}{n_0}\,\frac{dn_0(\mu)}{d\mu}\,d\mu
\end{equation}
which gives the mass spectrum.

With $\widetilde{H}(v_{\perp})\,
dv_{\perp}$ being the probability of finding the perpendicular velocity in
the interval 
$[v_{\perp},v_{\perp}+dv_{\perp}]$, one obtains for the event rate
\begin{equation}
\Gamma = D_\mathrm{s} w_0 \int r_\mathrm{E}(x) H(x) v_{\perp} \widetilde{H}(v_{\perp})
\frac{dn_0(\mu)}{d\mu}\,d\mu\,dv_{\perp}\,dx
\end{equation}
or 
\begin{equation}
\Gamma = \frac{D_\mathrm{s} w_0}{<\!\!M\!\!>} \int r_\mathrm{E}(x) \rho(x) v_{\perp} 
\widetilde{H}(v_{\perp})\,\omega(\mu)\,d\mu
\,dv_{\perp}\,dx
\end{equation}
with the average mass $<\!\!M\!\!> = \frac{\rho_0}{n_0}$.

Let $v_\mathrm{c}$ be a characteristic velocity and $\zeta = \frac{v_{\perp}}{v_\mathrm{c}}$.
The probability density for $\zeta$ is then given by
\begin{equation}
\widetilde{K}(\zeta) = \widetilde{H}(\zeta v_\mathrm{c})\,v_\mathrm{c}\,.
\end{equation}
Note that $\widetilde{K}$ may depend on $x$. For any $x$, $\widetilde{K}$
is normalized as
\begin{equation}
\int \widetilde{K}(\zeta;x)\,d\zeta = 1\,.
\end{equation}
With these definitions and $r_\mathrm{E}^2 = r_0^2 \mu x(1-x)$, one gets
\begin{eqnarray}
\Gamma = D_\mathrm{s} w_0\,r_0\,v_\mathrm{c} \int \sqrt{\mu x(1-x)}\, H(x)\, \zeta\,\widetilde{K}(\zeta)
\,\cdot \nonumber \\
\cdot\,\frac{dn_0(\mu)}{d\mu}\,d\mu\,d\zeta\,dx\,.
\label{eq:rate0}
\end{eqnarray}

\section{The probability density for the mass}
\label{physparoneevent}

Eq.~(\ref{eq:rate0}) gives the total event rate which includes events
with all possible 
timescales from the mass spectrum and the distribution of the lens position
and velocity.
By adding an integration over $t_\mathrm{E}$ and a $\delta$-function 
one gets
\begin{eqnarray}
\Gamma = D_\mathrm{s} w_0\,r_0\,v_\mathrm{c} \int \sqrt{\mu x(1-x)}\, H(x)\, \zeta\,\widetilde{K}(\zeta)
\frac{dn_0(\mu)}{d\mu}\,\cdot \nonumber \\
\cdot \,\delta\left(t_\mathrm{E} - \frac{r_0}{v_\mathrm{c}\,\zeta}
\sqrt{\mu x(1-x)}\,\right)\,dt_\mathrm{E}\,d\mu\,d\zeta\,dx\,.
\end{eqnarray}
The event rate contribution for timescales $t_\mathrm{E}$ in the interval 
$[t_\mathrm{E},t_\mathrm{E}+dt_\mathrm{E}]$ is given by
$\frac{d\Gamma}{dt_\mathrm{E}}\,dt_\mathrm{E}$. 

Let us now compare different mass spectra which have only the mass $\mu'$,
i.e. 
\begin{equation}
\frac{dn_0}{d\mu}(\mu) \propto \delta(\mu-\mu')\,.
\end{equation}
If one assigns the same probability to any mass a-priori, one has
$\omega(\mu) = \delta(\mu-\mu')$,
i.e. 
\begin{equation}
\frac{dn_0}{d\mu}(\mu) = \frac{\rho_0}{\mu M_{\sun}}\,
\delta(\mu-\mu')\,.
\end{equation}

More generally, one can use any explicit form of the mass spectrum,
e.g. a power law for 
$\frac{dn_0}{d\mu}$ by using a weighting factor $\alpha\,\mu^p$, i.e.
\begin{equation}
\frac{dn_0}{d\mu}(\mu) = \alpha\,\mu^p\,\delta(\mu-\mu')\,,
\end{equation}
so that the case above corresponds to $p = -1$.

For the power-law mass spectra, one obtains
\begin{eqnarray}
\frac{d\Gamma}{dt_\mathrm{E}} = 
D_\mathrm{s} w_0\,r_0\,v_\mathrm{c}\,\alpha\,
\int\,{\mu'}^{p+1/2}\,\sqrt{x(1-x)}\, H(x)\,\cdot\nonumber \\ \cdot\, \zeta\,\widetilde{K}(\zeta)
\,\delta\left(t_\mathrm{E} - \frac{r_0}{v_\mathrm{c}\,\zeta}
\sqrt{\mu' x(1-x)}\,\right)\,d\zeta\,dx\,.
\end{eqnarray}

For a given $\mu'$, the probability for a timescale in the interval
$[t_\mathrm{E}, t_\mathrm{E}+dt_\mathrm{E}]$ is given by $\frac{1}{\Gamma}
\frac{d\Gamma}{dt_\mathrm{E}}\,dt_\mathrm{E}$ 
as a function of $t_\mathrm{E}$. This fact has been
used by GRI and RJM
to compare the distribution of the timescales $t_\mathrm{E}$ for
different masses $\mu'$. 
By exchanging the roles of $\mu'$ and $t_\mathrm{E}$ one obtains the contribution
of masses in the interval $[\mu',\mu'+d\mu']$ for events with $t_\mathrm{E}$ to
the event rate as ($\mu'$ is called $\mu$ in the following)
\begin{eqnarray}
\frac{d\Gamma}{d\mu} = D_\mathrm{s} w_0 r_0\,v_\mathrm{c}\,\alpha\,
\int \mu^{p+1/2}\,\sqrt{x(1-x)}\,H(x)\,\cdot \nonumber \\ \cdot \,\zeta\,
\widetilde{K}(\zeta)\,
\delta\left(\mu - \frac{t_\mathrm{E}^2 v_\mathrm{c}^2 \zeta^2}{
r_0^2 x(1-x)}\right) dx\,d\zeta\,,
\end{eqnarray}
so that $\frac{1}{\Gamma}\frac{d\Gamma}{d\mu}$ gives the
probability density for the mass $\mu$.

The normalization factor is obtained by integration over $\mu$, which gives
\begin{eqnarray}
\Gamma & = & D_\mathrm{s} w_0\,t_\mathrm{E}^{2p+1}\,v_\mathrm{c}^{2p+2}\,
r_0^{-2p} \alpha\,\cdot \nonumber \\
& & \quad \cdot\,\int [x(1-x)]^{-p}\,H(x)\,\zeta^{2p+2}\,
\widetilde{K}(\zeta)\,dx\,d\zeta \nonumber \\
& =  & D_\mathrm{s} w_0\,t_\mathrm{E}^{2p+1}\,v_\mathrm{c}^{2p+2}\,r_0^{-2p} 
\alpha \, T(-p,2p+2)\,,
\end{eqnarray}
where
\begin{equation}
T(r,s)  =  \int \left[x(1-x)\right]^r\,H(x)\, 
\zeta^s\,\widetilde{K}(\zeta)\,d\zeta\,dx\,. \label{eqt}
\end{equation}
For the case that the velocity distribution does not
depend on $x$, the function $T(r,s)$ separates as
\begin{equation}
T(r,s) = \Xi(r)\,W(s)\,,
\end{equation}
where
\begin{eqnarray}
\Xi(r) & =  & \int \left[x(1-x)\right]^r\,H(x)\,dx\,, \label{eqxi} \\ 
W(s) & = &  \int \zeta^s\,\widetilde{K}(\zeta)\,d\zeta\,. \label{eqw} 
\end{eqnarray}
The probability density for $\mu$ follows as
\begin{eqnarray}
\frac{1}{\Gamma}\,\frac{d\Gamma}{d\mu}  =  
\frac{1}{T(-p,2p+2)}\,
\left(\frac{r_0}{t_\mathrm{E}\,v_\mathrm{c}}\right)^{2p+1}\,
\int \mu^{p+1/2}\,\cdot\nonumber \\ \cdot\,
\sqrt{x(1-x)}\,H(x)\,\zeta\,\widetilde{K}(\zeta)\,
\delta\left(\mu - \frac{t_\mathrm{E}^2 v_\mathrm{c}^2 \zeta^2}{
r_0^2 x(1-x)}\right)\, dx\,d\zeta\,.
\end{eqnarray}
Note that the width $w_0$ has cancelled out. This is due to the fact that
the fit parameters are kept fixed and only the unknown quantities $\mu$,
$x$, and $v_{\perp}$ are varied. Implicitly, the same probability is 
assigned to each parameter for the different values of $\mu$, $x$, and
$\zeta$.

\section{Moments of the probability distributions for physical quantities}
The expectation value $<\!\!\mu\!\!>$ follows as
\begin{eqnarray}
<\!\!\mu\!\!>  & =  & \int \frac{1}{\Gamma}\,\frac{d\Gamma}{d\mu}\,
\mu\,d\mu \nonumber \\ 
 & = & \frac{1}{T(-p,2p+2)}\, 
\int \mu(t_\mathrm{E},x,\zeta)\,
[x(1-x)]^{-p}\,H(x)\,\cdot \nonumber \\ & & \quad\cdot\,
\zeta^{2p+2}\,\widetilde{K}(\zeta)\,d\zeta\,dx \,,
\end{eqnarray}
and for a general quantity $G = G(t_\mathrm{E},x,\zeta)$ one obtains
\begin{eqnarray}
<\!\!G\!\!> & = & \int \frac{1}{\Gamma}\,\frac{d\Gamma}{d\mu}\,
G\,d\mu  \nonumber \\
 & = & \frac{1}{T(-p,2p+2)}\, 
\int G(t_\mathrm{E},x,\zeta)\,
[x(1-x)]^{-p}\,H(x)\,\cdot \nonumber \\
& & \quad \cdot\,
\zeta^{2p+2}\,\widetilde{K}(\zeta)\,d\zeta\,dx \,.
\end{eqnarray}

This means that one averages over $x$ with the density function
$[x(1-x)]^{-p}\,H(x)$ and over $\zeta$ with the density function
$\zeta^{2p+2}\,\widetilde{K}(\zeta)$.

For the quantity $G$ being of the form
\begin{equation}
G(t_\mathrm{E},x,\zeta) = G_0(t_\mathrm{E})\,[x(1-x)]^k\,\zeta^l
\end{equation}
one obtains
\begin{equation}
<\!\!G\!\!> = G_0\,\frac{T(k-p,l+2p+2)}{T(-p,2p+2)}
 = G_0\,F(p,k,l)\,.
\end{equation}
For the mass, one has
\begin{equation}
\mu(t_\mathrm{E},x,\zeta) = \frac{t_\mathrm{E}^2\,v_\mathrm{c}^2}{r_0^2}\,
\frac{\zeta^2}{x(1-x)}\,,
\end{equation}
so that $k=-1$ and $l=2$, and
\begin{equation}
<\!\!\mu\!\!> = \frac{t_\mathrm{E}^2\,v_\mathrm{c}^2}{r_0^2}\,
\frac{T(-1-p,2p+4)}{T(-p,2p+2)}\,.
\end{equation}
For $p = -1$, this gives
\begin{equation}
<\!\!\mu\!\!> = \frac{t_\mathrm{E}^2\,v_\mathrm{c}^2}{r_0^2}\,
\frac{T(0,2)}{\Xi(1)}\,,
\end{equation}
which is the same value as for the average mass in the mass spectrum
(if one uses only the information from a single event),
which can obtained by the method of mass moments
described in Sect.~\ref{massmoments}. Note however that 
the probability density for $\mu$ for a specific event and
the mass spectrum are different quantities and 
that the higher moments are different.

If the velocity distribution does not depend on $x$, 
one gets for the expectation value of the mass
\begin{equation}
<\!\!\mu\!\!> = 
\frac{t_\mathrm{E}^2\,v_\mathrm{c}^2}{r_0^2}\,
\frac{\Xi(0)\,W(2)}{\Xi(1)}\,.
\end{equation}

$\Xi(0)$ is related to the surface mass density $\Sigma$ by
\begin{equation}
\Sigma = \int\limits_0^{D_\mathrm{s}} \rho(D_\mathrm{d})\,dD_\mathrm{d} 
= D_\mathrm{s} \rho_0 \int\limits_0^1 H(x)\,dx 
= D_\mathrm{s} \rho_0\,\Xi(0)\,,
\end{equation}
and $\Xi(1)$ is related 
to the optical depth $\tau$ by
\begin{eqnarray}
\tau & = & \int\limits_0^{D_\mathrm{s}} \frac{4\pi G D}{c^2}\,\rho(D_\mathrm{d})\,
dD_\mathrm{d} \nonumber \\
& = & \frac{4G}{c^2}\,D_\mathrm{s}^2\,\rho_0\,\pi \int\limits_0^1 H(x) x(1-x)\,dx \nonumber \\
& = & r_0^2\,\frac{D_\mathrm{s} \rho_0}{M_{\sun}}\,\pi\,\Xi(1) \,.
\end{eqnarray}
Using these results, and noting that 
\begin{equation}
<\!\!v_{\perp}^2\!\!> = v_\mathrm{c}^2\,W(2)\,,
\end{equation}
the expectation value of the mass can be written in the
convenient form
\begin{equation}
<\!\!\mu\!\!> 
= t_\mathrm{E}^2\,v_\mathrm{c}^2\,W(2)\,\frac{\pi}{\tau}
\,\frac{\Sigma}{M_{\odot}}
\end{equation}
using the optical depth $\tau$, the area number density of the lenses $\Sigma$,
the average square of the velocity
$<\!\!v_{\perp}^2\!\!>$, and the timescale $t_\mathrm{E}$.
Note that $<\!\!\mu\!\!>$ depends only on $<\!\!v_{\perp}^2\!\!>$,
not on the form of the velocity distribution.

To investigate the distribution of $G$ around $<\!\!G\!\!>$, I define
\begin{equation}
\kappa_G = \frac{G}{<\!\!G\!\!>} = 
\frac{[x(1-x)]^k\,\zeta^l}{F(p,k,l)}\,.
\end{equation}

Using 
\begin{equation}
\widetilde{\mu}(t_\mathrm{E},x,\zeta) = \mu\,\frac{r_0^2}{t_\mathrm{E}^2 v_\mathrm{c}^2}
 = \frac{\zeta^2}{x(1-x)}\,,
\end{equation}
one obtains for 
$\kappa_G$
the probability density
\begin{eqnarray}
p(\kappa_G) & = & \frac{1}{\Gamma}\,\frac{d\Gamma}{d\kappa_G} \nonumber \\
 & = & \frac{1}{T(-p,2p+2)}\,
\left(\frac{r_0}{t_\mathrm{E}\,v_\mathrm{c}}\right)^{2p+1}\,
\int [\mu(\kappa_G)]^{p+1/2}\,\cdot\nonumber \\ & & \quad \cdot\,
\sqrt{x(1-x)}\,H(x)\,\zeta\,
\widetilde{K}(\zeta)\,\cdot \nonumber \\
& & \quad \cdot\, \delta\left(\kappa_G - \kappa_G(x,\zeta)\right)\,
dx\,d\zeta \nonumber \\
 & = & \frac{1}{T(-p,2p+2)}\,
\int [\widetilde{\mu}(\kappa_G)]^{p+1/2}\,\sqrt{x(1-x)}\,\cdot \nonumber \\
& & \quad \cdot\,H(x)\,\zeta\,
\widetilde{K}(\zeta)\,\cdot \nonumber \\
& & \quad \cdot\, \delta\left(\kappa_G - \frac{[x(1-x)]^k \zeta^l}{
F(p,k,l)}\right)\,
dx\,d\zeta \nonumber \\
 & = & \frac{F(p,k,l)}{T(-p,2p+2)}\,\frac{1}{|l|}
\int [\widetilde{\mu}(\kappa_G)]^{p+1/2}\,\cdot \nonumber \nonumber
\\ & & \quad \cdot\,[x(1-x)]^{1/2-k}\,H(x)\,
\zeta^{-l+2}\,\widetilde{K}(\zeta)\,\cdot \nonumber \\
& & \quad \cdot\,\delta\left(\frac{\kappa_G\,F(p,k,l)}{[x(1-x)]^k}- \zeta^l\right)\,
dx\,d(\zeta^l) \nonumber \\
 & = & \frac{[F(p,k,l)]^{\frac{2p+3}{l}}}{|l|\,T(-p,2p+2)}\,
\kappa_G^{\frac{2p-l+3}{l}}\,\cdot \nonumber\\
& & \quad \cdot\,\int [x(1-x)]^{-p-(3+2p)\frac{k}{l}}\,H(x)\,\cdot\nonumber \\
& & \quad \cdot\,
\widetilde{K}\left(\left(\frac{\kappa_G\,F(p,k,l)}{[x(1-x)]^k}\right)
^{1/l}\right)\,
dx\,. \label{pkapggen}
\end{eqnarray}
{\em Note that this distribution depends only on the form of the distributions
of $x$ and $\zeta$ and not on any physical parameters like $v_\mathrm{c}$
or $\rho_0$.}

The distribution of $\lambda_G = \mathrm{lg}~\kappa_G$ is given by
\begin{eqnarray}
\psi(\lambda_G) & = & \frac{\ln 10 \cdot 
\left[F(k,p,l)\cdot 10^{\lambda_G}\right]^{\frac{2p+3}{l}}
}{|l|\, T(-p,2p+2)}\, \cdot \nonumber \\
& & \quad \cdot\,\int [x(1-x)]^{-p-(3+2p)\frac{k}{l}}\,H(x) \,\cdot \nonumber
\\ & & \quad \cdot\,
\widetilde{K}\left(\left(\frac{10^{\lambda_G}}{
[x(1-x)]^k}\,F(p,k,l)\right)^{1/l}\right)
dx \,. \label{psilamggen}
\end{eqnarray}

All moments of the probability distribution of $G$ can be reduced
to the function $T(k,j)$, which separates as the product of 
$\Xi(k)$ and $W(j)$ if 
the velocity distribution does not depend on $x$. 
For the $n$-th moment one has
\begin{equation}
<\!\!G^n\!\!> = \int \frac{1}{\Gamma}\,\frac{d\Gamma}{d\mu}\,G^n\,d\mu
= G_0^n\,F(p,nk,nl)\,.
\end{equation}
The relative deviation of $G$ is given by
\begin{equation}
\sigma_{\kappa_G} = \frac{\sigma_G}{<\!\!G\!\!>} = \frac{\sqrt{<\!\!G^2\!\!> - <\!\!G\!\!>^2}}{ 
<\!\!G\!\!>} =  \sqrt{\frac{<\!\!G^2\!\!>}{<\!\!G\!\!>^2} -1} \,.
\end{equation}
With
\begin{eqnarray}
F_2(p,k,l)  & = & \frac{F(p,2k,2l)}{[F(p,k,l)]^2} \nonumber \\
& =&
\frac{T(2k-p,2l+2p+2)}{[T(k-p,l+2p+2)]^2}\,T(-p,2p+2) \,,
\end{eqnarray}
one obtains
\begin{equation}
\sigma_{\kappa_G} = \frac{\sigma_G}{<\!\!G\!\!>} = \sqrt{F_2(p,k,l) - 1}\,.
\end{equation} 

\section{The method of mass moments}
\label{massmoments}
For the moments of the $t_\mathrm{E}$-distribution for several events, 
one obtains following
RJM
\begin{eqnarray}
\overline{t_\mathrm{E}^m} & = &
\frac{1}{\Gamma} 
\int \frac{d\Gamma}{dt_\mathrm{E}}\,t_\mathrm{E}^m\,dt_\mathrm{E} \nonumber \\
& = &
\frac{D_\mathrm{s} w_0\,r_0\,v_\mathrm{c}}{\Gamma}
\int \sqrt{\mu x(1-x)}\,H(x)\,\zeta\,\widetilde{K}(\zeta)
\frac{dn_0(\mu)}{d\mu}\,\cdot \nonumber \\ & & \quad\cdot\,t_\mathrm{E}^m\,
\delta\left(t_\mathrm{E} - \frac{r_0}{v_\mathrm{c}\,\zeta}
\sqrt{\mu x(1-x)}\,\right)\,dt_\mathrm{E}\,d\mu\,d\zeta\,dx \nonumber \\
& = &
\frac{D_\mathrm{s} w_0\,r_0^{m+1}\,v_\mathrm{c}^{1-m}}{\Gamma}
\int \mu^{\frac{m+1}{2}}\frac{dn_0(\mu)}{d\mu}\,d\mu\,\cdot\nonumber \\
& & \quad\cdot\,
\int \left[x(1-x)\right]^{\frac{m+1}{2}} H(x)\, 
\zeta^{1-m}\,\widetilde{K}(\zeta)\,d\zeta\,dx \nonumber \\
& = &
\frac{D_\mathrm{s} w_0\,r_0^{m+1}\,v_\mathrm{c}^{1-m}}{\Gamma}
n_0\;\overline{\mu^{\frac{m+1}{2}}}\;T\left(\frac{m+1}{2},1-m\right)
\,,
\end{eqnarray}
where 
\begin{equation}
\overline{\mu^k}  = \frac{1}{n_0} \int \mu^k\,\frac{dn_0(\mu)}{d\mu}\,d\mu
= \int \mu^k \omega(\mu)\,d\mu\,.
\end{equation}
$\overline{\mu^k}$ yields for the case $k=0$, $\overline{\mu^0} = 1$,
and for $k=1$ the average mass (in units of $M_{\sun}$) as
\begin{equation}
\overline{\mu} = \int \mu\, \omega(\mu)\,d\mu = \frac{1}{n_0}
\int \frac{dn_0(\mu)}{d\mu}\,\mu\,d\mu = \frac{\rho_0}{n_0\,M_{\sun}}\,.
\end{equation}
One sees that the moments of the mass spectrum can be extracted from the 
moments of the time-scale distribution if the spatial distribution of the 
mass density and the velocity distribution is given (RJM).

For the moments of the mass spectrum one gets
\begin{equation}
\overline{\mu^k}= \frac{\Gamma}{D_\mathrm{s} w_0\,n_0}\,
\frac{v_\mathrm{c}^{2k-2}}{r_0^{2k}}\,
\frac{1}{T(k,2-2k)}\,\overline{t_\mathrm{E}^{2k-1}}\,.
\end{equation}

This gives for $k=0$
\begin{equation}
1 = \overline{\mu^0}= \frac{\Gamma}{D_\mathrm{s} w_0\,n_0\,v_\mathrm{c}^2}\,
\frac{1}{T(0,2)}\,\overline{t_\mathrm{E}^{-1}}\,,
\end{equation}
which yields
\begin{equation}
n_0 = \frac{\Gamma}{D_\mathrm{s} w_0\,v_\mathrm{c}^2}\,
\frac{1}{T(0,2)}\,\overline{t_\mathrm{E}^{-1}}\,,
\end{equation}
so that
\begin{equation}
\overline{\mu^k} = \frac{v_\mathrm{c}^{2k}}{r_0^{2k}}\,
\frac{T(0,2)}{T(k,2-2k)}\,\frac{\overline{t_\mathrm{E}^{2k-1}}}
{\overline{t_\mathrm{E}^{-1}}}\,.
\end{equation}

For $k=1$, one obtains
\begin{equation}
\overline{\mu} = \frac{\overline{t_\mathrm{E}}}{\overline{t_\mathrm{E}^{-1}}}\,
\frac{v_\mathrm{c}^2}{r_0^2}\,\frac{T(0,2)}{T(1,0)} 
= \frac{\overline{t_\mathrm{E}}}{\overline{t_\mathrm{E}^{-1}}}\,
\frac{v_\mathrm{c}^2}{r_0^2}\,\frac{T(0,2)}{\Xi(1)}\,,
\end{equation}
since $T(1,0) = \Xi(1)$, which is the same value as obtained 
for the expectation value of the lens mass for $p = -1$
as derived above, if one considers a single observed event.

\section{Special forms of the velocity distribution 
$\widetilde{K}(\zeta)$}

\subsection{Maxwell distribution}
A distribution 
\begin{equation}
\widetilde{H}(v_\perp) = \frac{2 v_{\perp}}{v_\mathrm{c}^2}\,
\exp\left\{-\frac{v_{\perp}^2}{v_\mathrm{c}^2}\right\}
\end{equation}
implies
\begin{equation}
\widetilde{K}(\zeta) = 2\zeta \exp\left\{-\zeta^2\right\}\,,
\end{equation}
and for $p(\kappa_G)$ and $\psi(\lambda_G)$ one has
from Eqs.~(\ref{pkapggen}) and~(\ref{psilamggen})
\begin{eqnarray}
p(\kappa_G)  =  \frac{2 [F(p,k,l)]^{\frac{2p+4}{l}}}{|l|\,\Xi(-p) W(2p+2)}\,
\kappa_G^{\frac{2p+4}{l}-1}\,\cdot \nonumber \\
\cdot\,\int [x(1-x)]^{-p-(4+2p)\frac{k}{l}}\,H(x)\,\cdot \nonumber \\
\cdot\,
\exp\left\{-\left(\frac{\kappa_G\,F(p,k,l)}{[x(1-x)]^k}\right)
^{2/l}\right\}\,
dx\,, \\
\psi(\lambda_G)  =  \frac{2 \ln 10 \cdot 
\left[F(k,p,l)\cdot 10^{\lambda_G}\right]^{\frac{2p+4}{l}}}{
|l|\, \Xi(-p) W(2p+2)}\, \cdot \nonumber \\
\cdot\,\int [x(1-x)]^{-p-(4+2p)\frac{k}{l}}\,H(x)\,\cdot\nonumber \\
\cdot\,
\exp\left\{-\left(\frac{10^{\lambda_G}}{
[x(1-x)]^k}\,F(p,k,l)\right)^{2/l}\right\}
dx \,.
\end{eqnarray}
 
For $p=-1$ and separation of the spatial distribution and
velocity distribution, these equations read
\begin{eqnarray}
p(\kappa_G)  =  \frac{2}{|l|\,\Xi(1)}\,
\left(\frac{\Xi(k+1)W(l)}{\Xi(1)}\right)^{2/l}\,
\kappa_G^{\frac{2}{l}-1}\,\cdot \nonumber \\
 \cdot\,\int [x(1-x)]^{1-2\frac{k}{l}}\,H(x)\,\cdot\nonumber \\
\quad \cdot\,
\exp\left\{-\left(\frac{\kappa_G}{[x(1-x)]^k}\,
\frac{\Xi(k+1)W(l)}{\Xi(1)}\right)
^{2/l}\right\}\,
dx\,, 
 \label{ppsimasseq1} \\
\psi(\lambda_G)  =  \frac{2 \ln 10 \cdot 
 10^{\frac{2 \lambda_G}{l}}}{
l|\, \Xi(1)}\,
\left(\frac{\Xi(k+1)W(l)}{\Xi(1)}\right)^{2/l} \cdot \nonumber \\
\quad \cdot\,\int [x(1-x)]^{1-2\frac{k}{l}}\,H(x) \,\cdot\nonumber \\
\quad \cdot\,
\exp\left\{-\left(\frac{10^{\lambda_G}}{
[x(1-x)]^k}\,\frac{\Xi(k+1)W(l)}{\Xi(1)}\right)^{2/l}\right\}
dx  \,.
\label{ppsimasseq2}
\end{eqnarray}
Examples for these distributions are shown in Sect.~\ref{estpar:pkhalo},
where the galactic halo is discussed.

\subsection{Fixed velocity}
A model with a fixed velocity corresponds to
\begin{equation}
\widetilde{K}(\zeta) = \delta(\zeta-1)\,.
\end{equation}
It follows that $W(n) = 1\; \forall n$ and $F(p,k,l)$ does not depend on
$l$, so that
\begin{equation}
F(p,k,l) = \Phi(p,k)
\end{equation}
with
\begin{equation}
\Phi(p,k) = \frac{\Xi(k-p)}{\Xi(-p)}\,.
\end{equation}

With Eq.~(\ref{pkapggen}), $p(\kappa_G)$ follows as
\begin{eqnarray}
p(\kappa_G) & = & \frac{[\Phi(p,k)]^{\frac{2p+3}{l}}}{|l|\,\Xi(-p)}\,
\kappa_G^{\frac{2p+3}{l} - 1}\,\int [x(1-x)]^{-p-(3+2p)\frac{k}{l}}\,\cdot\nonumber \\
& & \quad\cdot\,
H(x)\,\delta\left(\left(\frac{\kappa_G\,\Phi(p,k)}{[x(1-x)]^k}\right)^{1/l}
-1\right)\,dx \nonumber\\
& = & \frac{[\Phi(p,k)]^{\frac{2p+3}{l}}}{|l|\,\Xi(-p)}\,
\kappa_G^{\frac{2p+3}{l} - 1}\,\int [x(1-x)]^{-p-(2+2p)\frac{k}{l}}\,\cdot\nonumber \\
& & \quad\cdot\,
H(x)\,\cdot\nonumber \\
& & \quad\cdot
\,\delta\left(\left(\kappa_G\,\Phi(p,k)\right)^{1/l}
-[x(1-x)]^{k/l}\right)\,dx\,.
\end{eqnarray}
With the substitution
\begin{equation}
d[x(1-x)]^{k/l} = \frac{k}{l}\,[x(1-x)]^{k/l-1}\,(1-2x)\,dx\,,
\end{equation}
one gets
\begin{eqnarray}
p(\kappa_G)   & = &\frac{[\Phi(p,k)]^{\frac{2p+3}{l}}}{|k|\,\Xi(-p)}\,
\kappa_G^{\frac{2p+3}{l} - 1}\,\cdot\nonumber \\
 & & \quad\cdot\,\int [x(1-x)]^{-p+1-(3+2p)\frac{k}{l}}
\,
\left|\frac{1}{1-2x}\right|\,H(x)\,\cdot \nonumber\\
& & \quad \cdot\,
\delta\left(\left(\kappa_G\,\Phi(p,k)\right)^{1/l}
-[x(1-x)]^{k/l}\right) \,\cdot\nonumber \\
& & \quad\cdot\,d\left([x(1-x)]^{k/l}\right) \,.
\end{eqnarray}

The argument of the $\delta$-function is zero for
\begin{equation}
 x = \frac{1}{2}\left(1 \pm \sqrt{1 -
4\left(\kappa_G\,\Phi(p,k)\right)^{1/k}}\,\right)\,,
\end{equation}
so that
\begin{equation}
1-2x = \mp \sqrt{1 -
4\left(\kappa_G\,\Phi(p,k)\right)^{1/k}}\,.
\end{equation}

This yields for $p(\kappa_G)$ 
\begin{eqnarray}
p(\kappa_G)  & = & \frac{[\Phi(p,k)]^{\frac{1-p}{k}}}{|k|\,\Xi(-p)}\,
\kappa_G^{\frac{1-p}{k} - 1}\,
\frac{1}{\sqrt{1 -
4\left(\kappa_G\,\Phi(p,k)\right)^{1/k}}}\,\cdot \nonumber\\
& & \cdot\,
\left[H\left(\frac{1}{2}\left(1 + \sqrt{1 -
4\left(\kappa_G\,\Phi(p,k)\right)^{1/k}}\,\right)\right)
\,+\nonumber\right. \\
& & +\, \left.H\left(\frac{1}{2}\left(1 - \sqrt{1 -
4\left(\kappa_G\,\Phi(p,k)\right)^{1/k}}\,\right)\right)\right] \,.
\end{eqnarray}
Note that $p(\kappa_G)$ does not depend on $l$.

Since there exist (real) solutions for $x$ only for
\begin{equation}
\left[\kappa_G\,\Phi(p,k)\right]^{1/k} \leq \frac{1}{4}\,,
\end{equation}
$\kappa_G$ is restricted by
$\kappa_G \leq \kappa_{G,\mathrm{crit}}$ for $k > 0$,
or by
$\kappa_G \geq \kappa_{G,\mathrm{crit}}$ for $k < 0$,
where
\begin{equation}
\kappa_{G,\mathrm{crit}} = \frac{1}{4^k\,\Phi(p,k)}\,.
\end{equation}
Since
$<\!\!G\!\!> = G_0\,\Phi(p,k)$,
the critical value of $G$ is
\begin{equation}
G_\mathrm{crit} = \kappa_{G,\mathrm{crit}}\,<\!\!G\!\!> = \frac{G_0}{4^k}\,,
\end{equation}
which is a maximum for $k>0$ and a minimum for $k<0$.

The distribution of $G$ can be written in terms of
\begin{equation}
\widetilde{\kappa_G} = \frac{G}{G_\mathrm{crit}} = \frac{\kappa_G}{ 
\kappa_{G,\mathrm{crit}}} = \kappa_G\,4^k\,\Phi(p,k)\,,
\end{equation}
which yields the probability density
\begin{eqnarray}
\widetilde{p}(\widetilde{\kappa_G})  & = & \frac{1}{4^{1-p} |k|\,\Xi(-p)}\,
\widetilde{\kappa_G}^{\frac{1-p}{k} - 1}\,
\frac{1}{\sqrt{1 - \widetilde{\kappa_G}^{1/k}}}\,\cdot \\
& & \quad \cdot\,
\left[H\left(\frac{1}{2}\left(1 + \sqrt{1 -
\widetilde{\kappa_G}^{1/k}}\,\right)\right)
\,+\nonumber\right. \\
& & \quad+\, \left.H\left(\frac{1}{2}\left(1 - \sqrt{1 -
\widetilde{\kappa_G}^{1/k}}\,\right)\right)\right] \,,
\end{eqnarray}
and for $\widetilde{\lambda_G} = \mathrm{lg}~\widetilde{\kappa_G}$ one
gets the probability density
\begin{eqnarray}
\widetilde{\psi}(\widetilde{\lambda_G})  & = & \frac{\ln 10}{
4^{1-p} |k|\,\Xi(-p)}\,
10^{\widetilde{\lambda_G}\,\frac{1-p}{k}}\,
\frac{1}{\sqrt{1 - 10^{\widetilde{\lambda_G}/k}}}\,\cdot \\
& & \quad \cdot\,
\left[H\left(\frac{1}{2}\left(1 + \sqrt{1 -
10^{\widetilde{\lambda_G}/k}}\,\right)\right)
\,+ \nonumber\right. \\
& & \quad +\,\left.H\left(\frac{1}{2}\left(1 - \sqrt{1 -
10^{\widetilde{\lambda_G}/k}}\,\right)\right)\right] \,,
\end{eqnarray}
Since
\begin{eqnarray}
\lambda_G & = & \mathrm{lg}~\kappa_G = \mathrm{lg}~\left(\widetilde{\kappa_G}\,
\kappa_{G,\mathrm{crit}}\right) \nonumber \\
& = & \widetilde{\lambda_G}
- \mathrm{lg}~\Phi(p,k) - 2k\,\mathrm{lg}~2\,,
\end{eqnarray}
$\psi(\lambda_G)$ is given by
\begin{equation}
\psi(\lambda_G) = \widetilde{\psi}(\lambda_G+\mathrm{lg}~\Phi(p,k)+2k\,\mathrm{lg}~2)\,.
\end{equation}

\section{Estimates and probability distributions for
physical quantities for a simple galactic halo model}
\label{estpar:certain}
\label{estpar:pkhalo}
The quantities to be discussed here are
\begin{itemize}
\item the transverse velocity $v_{\perp}$,
\item the Einstein radius $r_\mathrm{E}$ (which immediately yields
the projected distance $2 \chi r_\mathrm{E}$),
\item the mass $M = \mu M_{\sun}$, 
\item the rotation period $T$ of a binary. 
\end{itemize}

For these quantities, $G(t_\mathrm{E},x,\zeta)$, $k$, $l$, and $F(-1,k,l)$
ares shown in Table~\ref{estpartabgen}. The right expression for $F(-1,k,l)$
is valid if the velocity distribution does not depend on $x$, $\rho$ denotes
the semimajor axis in units of Einstein radii.

\begin{table*}
\caption{$G(t_\mathrm{E},x,\zeta)$, $k$, $l$, and $F(-1,k,l)$ for 
$v_{\perp}$, $r_\mathrm{E}$, $\mu$, $T$} 
\begin{flushleft}
\begin{tabular}{ccccc}
\hline\noalign{\smallskip}
$G$ & $G(t_\mathrm{E},x,\zeta)$ & $k$ & $l$ & $F(-1,k,l)$ \\ 
\noalign{\smallskip}\hline \noalign{\smallskip}
$v_{\perp}$ & $v_\mathrm{c}\,\zeta$ & 0 & 1 & $\frac{T(1,1)}{\Xi(1)} = W(1)$ \\ 
$r_\mathrm{E}$ & $t_\mathrm{E}\,v_\mathrm{c}\,\zeta$ &
0 & 1 & $\frac{T(1,1)}{\Xi(1)} = W(1)$ \\ 
$\mu$ & $\frac{t_\mathrm{E}^2 v_\mathrm{c}^2}{r_0^2}\,\frac{\zeta^2}{x(1-x)}$ &
-1 & 2 & $\frac{T(0,2)}{\Xi(1)} = \frac{\Xi(0) W(2)}{\Xi(1)}$ \\
$T$ & $\frac{4\pi}{c}\,\sqrt{\rho^3\,t_\mathrm{E}\,D_\mathrm{s}\,v_\mathrm{c}} \,
\sqrt{x(1-x)}\,\sqrt{\zeta}$ &
$\frac{1}{2}$ & $\frac{1}{2}$ & 
$\frac{T(\frac{3}{2},\frac{1}{2})}{\Xi(1)} = 
\frac{\Xi(\frac{3}{2}) W(\frac{1}{2})}{\Xi(1)}$ \\
\noalign{\smallskip}\hline
\end{tabular}
\end{flushleft}
\label{estpartabgen}
\end{table*}

For the galactic halo, a velocity distribution of
\begin{equation}
\widetilde{H}(v_{\perp}) = \frac{2 v_{\perp}}{v_c^2}\,\exp\left\{-
\frac{v_{\perp}^2}{v_\mathrm{c}^2}\right\}
\end{equation}
can be used, where ${\overline{v_{\perp}^2}} = v_\mathrm{c}^2$.

The mass density of halo objects is modelled as
\begin{equation}
\rho(r) = \rho_0\,\frac{a^2 + R_\mathrm{GC}^2}{a^2 + r^2}\,,
\end{equation}
where $r$ measures the distance from the Galactic center, $R_\mathrm{GC}$ is
the distance from the sun to the Galactic center, $a$ is a characteristic
core radius and $\rho_0$ is the local density at the position of the sun.

With the distance parameter $x$, which
measures the distance along the line-of-sight from the observer to the
LMC in units of the total distance $D_\mathrm{s}$, one obtains
\begin{equation}
\rho(x) = \rho_0\,\frac{ {1 + \frac{a^2}{R_\mathrm{GC}^2}}}{
 {1 + \frac{a^2}{R_\mathrm{GC}^2}} + x^2 \frac{D_\mathrm{s}}{R_\mathrm{GC}^2}
-2 x \frac{D_\mathrm{s}}{R_\mathrm{GC}} \cos \alpha}\,,
\end{equation}
where $\alpha$ is the angle between direction of the Galactic center and 
the direction of the LMC measured from the observer.

With 
\begin{equation}
\xi_\mathrm{s} = \frac{D_\mathrm{s}}{R_\mathrm{GC}}\,, \quad
A = 1+\frac{a^2}{R_\mathrm{GC}^2} \,, \quad
B = -2 \cos \alpha\,,
\end{equation}
H(x) can be written as
\begin{equation}
H(x) = \frac{A}{A+Bx\xi_\mathrm{s}+x^2 \xi_\mathrm{s}^2}\,.
\end{equation}

Let the halo be extended to a distance of $D_\mathrm{h}$ along the line-of-sight.
With $\xi_h = D_\mathrm{h}/R_\mathrm{GC}$ and $\xi = \xi_\mathrm{h}/\xi_\mathrm{s}$,
one obtains
\begin{eqnarray}
\Xi(0) & = & \int\limits_0^{\xi} H(x)\,dx \nonumber\\
& = & \int\limits_0^{\xi}  \frac{A}{A+Bx\xi_\mathrm{s}+x^2 \xi_\mathrm{s}^2}\,dx \nonumber\\
& = & \frac{A}{\xi_\mathrm{s}} \int\limits_0^{\xi_\mathrm{h}}  
\frac{1}{A+B\widetilde{x}+\widetilde{x}^2 }\,d\widetilde{x} \nonumber \\
& = & \frac{A}{\xi_\mathrm{s}}\,\frac{2}{\sqrt{4A-B^2}}\,
\left[\arctan \frac{2 \xi_\mathrm{h} + B}{\sqrt{4A-B^2}}\right.\, - \nonumber \\
& & \quad - \, \left.
\arctan \frac{B}{\sqrt{4A-B^2}}\right]\,,
\end{eqnarray}
and
\begin{eqnarray}
\Xi(1) & = &  \int\limits_0^{\xi} x(1-x)\,H(x)\,dx \nonumber \\
& = & A \int\limits_0^{\xi}  \frac{x (1-x)}{A+Bx\xi_\mathrm{s}+x^2 \xi_\mathrm{s}^2}\,dx \nonumber \\
& = & \frac{A}{\xi_\mathrm{s}^3} \int\limits_0^{\xi_\mathrm{h}}  
\frac{\widetilde{x}(\xi_\mathrm{s}-\widetilde{x})}{A+B\widetilde{x}+\widetilde{x}^2 }\,d\widetilde{x} \nonumber\\
& = &  \frac{A}{\xi_\mathrm{s}^3}\,\left\{-\xi_\mathrm{h}
+ \frac{1}{2}(\xi_\mathrm{s} + B) \ln{\frac{A+B\xi_\mathrm{h}+\xi_\mathrm{h}^2}{A}} - 
\right.\nonumber\\
& & \quad
\left.-\, \frac{\xi_\mathrm{s}B+B^2-2A}{\sqrt{4A-B^2}}\,\left[
\arctan \frac{2 \xi_\mathrm{h} + B}{\sqrt{4A-B^2}} \right.\right.\,-\nonumber \\
& & \quad - \, \left.\left.
\arctan \frac{B}{\sqrt{4A-B^2}}\right]\right\}\,.
\end{eqnarray}
Other values of $\Xi(r)$ can be obtained numerically.

Selected values for $\Xi(r)$ and $W(s)$ and $F(-1,k,l)$ are shown in 
Tables~\ref{halo:Xi}, \ref{halo:W}, and~\ref{halo:F}, using
the values used by Paczy{\'n}ski (\cite{paczynski1})\footnote{Using slightly
different values for $D_\mathrm{s}$ and $R_\mathrm{GC}$, and varying the 
core radius $a$ between 0 and 8~kpc yields estimates which differ
by about~5~\%.}
\begin{eqnarray}
D_\mathrm{s} = 50~\mathrm{kpc}\,,\quad
R_\mathrm{GC} = 10~\mathrm{kpc}\,,\quad
x_\mathrm{s} = 5\,,\quad
x_\mathrm{h} = 5\,,\quad \nonumber \\
\alpha = 82^{\circ}\,,\quad
a = 0\,. \nonumber
\end{eqnarray}

\begin{table}[htbp]
\caption{Selected values for $\Xi(r)$} 
\begin{flushleft}
\begin{tabular}{cc}
\hline\noalign{\smallskip}
$r$ & $\Xi(r)$ \\ \noalign{\smallskip}\hline\noalign{\smallskip}
0 & 0.305 \\
$\frac{1}{2}$ & 0.105 \\
1 & 0.0407 \\
$\frac{3}{2}$ & 0.0168 \\
2 & 0.00721 \\
\noalign{\smallskip}\hline
\end{tabular}
\end{flushleft}
\label{halo:Xi}
\end{table}

\begin{table}[htbp]
\caption{Selected values for $W(s)$} 
\begin{flushleft}
\begin{tabular}{cc}
\hline\noalign{\smallskip}
$s$ & $W(s)$ \\ \noalign{\smallskip}\hline\noalign{\smallskip} 
$-\frac{1}{2}$ & 1.225 \\ 
0 & 1.000 \\ 
$\frac{1}{2}$ & 0.906 \\ 
1 & 0.886 \\ 
$\frac{3}{2}$ & 0.919 \\ 
2 & 1.000 \\ \noalign{\smallskip}\hline
\end{tabular}
\end{flushleft}
\label{halo:W}
\end{table}

\begin{table}[htbp]
\caption{$F(-1,k,l)$ for $v_{\perp}$, $r_\mathrm{E}$, $\mu$, $T$} 
\begin{flushleft}
\begin{tabular}{cc}
\hline\noalign{\smallskip}
$G$ & $F(-1,k,l)$ \\ \noalign{\smallskip}\hline\noalign{\smallskip} 
$v_{\perp}$,$r_\mathrm{E}$ & 0.886 \\ 
$\mu$ & 7.49 \\ 
$T$ & 0.374 \\ \noalign{\smallskip}\hline
\end{tabular}
\end{flushleft}
\label{halo:F}
\end{table}

From these values, one obtains for the expectation values:
\begin{eqnarray}
<\!\!v_{\perp}\!\!> & = & \frac{1}{2}\sqrt{\pi}\,{v_\mathrm{c}} = 0.886\,{v_\mathrm{c}}\,,
\label{haloestfirst} \\
<\!\!r_\mathrm{E}\!\!> & = & 0.107\,\left(\frac{v_\mathrm{c}}{210~\mathrm{km/s}}\right)\,
\,\left(\frac{t_\mathrm{E}}{1~\mathrm{d}}\right)\,\mathrm{AU}\,, \\
<\!\!\mu\!\!> & = & 2.71\cdot 10^{-4}\,\left(\frac{v_\mathrm{c}}{210~\mathrm{km/s}}\right)^2\,
\,\left(\frac{t_\mathrm{E}}{1~\mathrm{d}}\right)^2\,, \\
<\!\!T\!\!> & = & 2.63\,\rho^{3/2}\,\left(\frac{v_\mathrm{c}}{210~\mathrm{km/s}}\right)^{1/2}\,
\,\left(\frac{t_\mathrm{E}}{1~\mathrm{d}}\right)^{1/2}\,\mathrm{a}\,. 
 \label{haloestlast} 
\end{eqnarray}

With $\rho_0 = \frac{v_\mathrm{c}^2}{4 \pi G R_\mathrm{GC}^2}$ one obtains
\begin{equation}
\rho_0 = 8.16\cdot 10^{-3}\,
\left(\frac{v_\mathrm{c}}{210~\mathrm{km/s}}\right)^2\,
\frac{M_{\sun}}{\mathrm{pc}^3}\,,
\end{equation}
so that for $v_\mathrm{c} = 210~\mathrm{km/s}$, one gets
$\Sigma = 124\,{M_{\sun}}/{\mathrm{pc}^2}$
and
$\tau = 5\cdot 10^{-7}$.

The distribution of $\kappa_{\mu} = \mu/\!<\!\!\mu\!\!>$ is given 
by 
\begin{eqnarray}
p(\kappa_{\mu}) = \frac{\Xi(0)}{(\Xi(1))^2} \int
x^2(1-x)^2\,H(x)\,\cdot\nonumber \\
\cdot\,\exp\left\{-\frac{\Xi(0)}{\Xi(1)}\,\kappa_{\mu} x(1-x)\right\}\,
dx\,.
\label{pkapmu}
\end{eqnarray}
and the probability density $\psi(\lambda_{\mu})$
is
\begin{eqnarray}
\psi(\lambda_{\mu}) = \frac{\Xi(0)}{(\Xi(1))^2}\,10^{\lambda_{\mu}}\,{\ln 10} \int
x^2(1-x)^2\,H(x)\,\cdot\nonumber \\
\cdot\,\exp\left\{-\frac{\Xi(0)}{\Xi(1)}\,10^{\lambda_{\mu}} x(1-x)\right\}\,
dx\,.
\end{eqnarray}
From Eq.~(\ref{pkapmu}) the probability density $P(\mu)$ for the mass
$\mu$ (in units of $M_{\sun}$) follows as
\begin{eqnarray}
P(\mu) = \frac{r_0^2}{t_\mathrm{E}^2\,v_\mathrm{c}^2}\,
\frac{1}{\Xi(1)}\,\int
x^2(1-x)^2\,H(x)\,\cdot\nonumber \\
\cdot\,\exp\left\{-\frac{\mu\,r_0^2\,x(1-x)}{t_\mathrm{E}^2\,
v_\mathrm{c}^2}\right\}\,dx\,,
\end{eqnarray}
which differs from the probability given by Jetzer \& Mass{\'o} (\cite{JM}, 
Eq.~(8)),
by a factor
$\mu^2/t_\mathrm{E}^2$. Note that this probability density has to be of the form
\begin{equation}
P(\mu) = \frac{1}{t_\mathrm{E}^2}\,f\left(\frac{\mu}{t_\mathrm{E}^2}\right)\,,
\end{equation}
since $\mu \propto t_\mathrm{E}^2$, to ensure normalization for any $t_\mathrm{E}$.

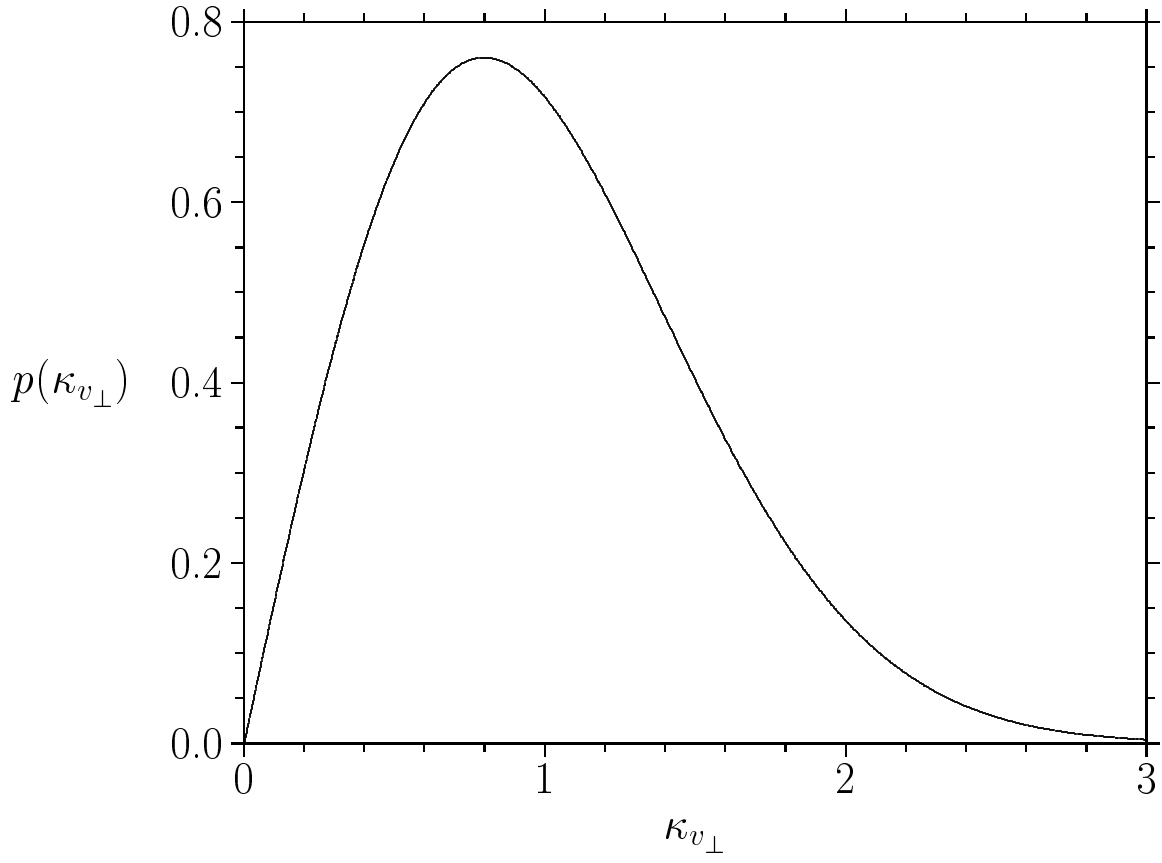
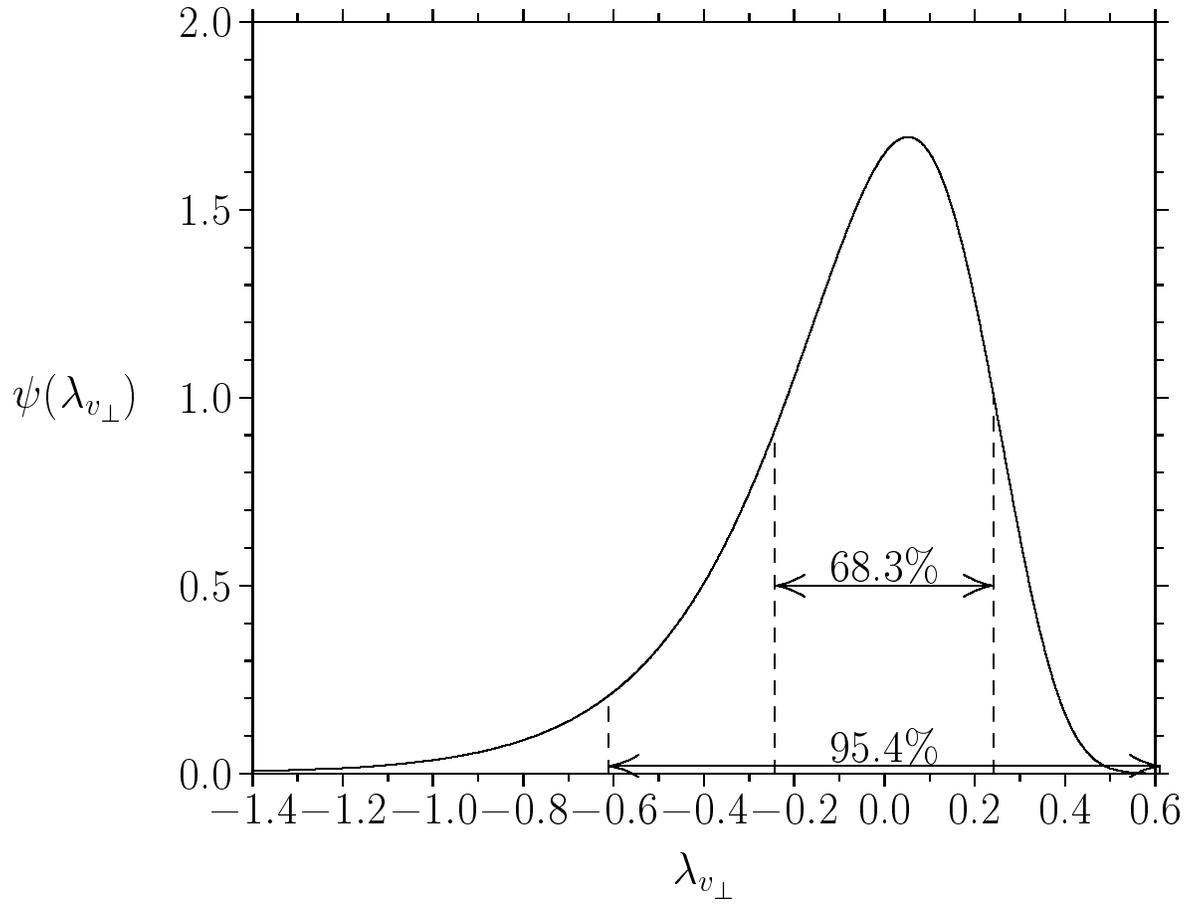
\begin{figure*}
\vspace{7cm}
\caption{The probability density $p(\kappa_{v_{\perp}}) = 
p(\kappa_{r_\mathrm{E}})$ 
for $\kappa_{v_{\perp}} = v_{\perp} /\!<\!\!v_{\perp}\!\!> = 
r_\mathrm{E} /<\!\!r_\mathrm{E}\!\!> = \kappa_{r_\mathrm{E}}$ (left) and
The probability density $\psi(\lambda_{v_{\perp}}) = 
\psi(\lambda_{r_\mathrm{E}})$ 
with symmetric 68.3~\% and 95.4~\% intervals around $<\!\!v_{\perp}\!\!>$ or
$<\!\!r_\mathrm{E}\!\!>$ (right)}
\label{halo:pkre}
\end{figure*}

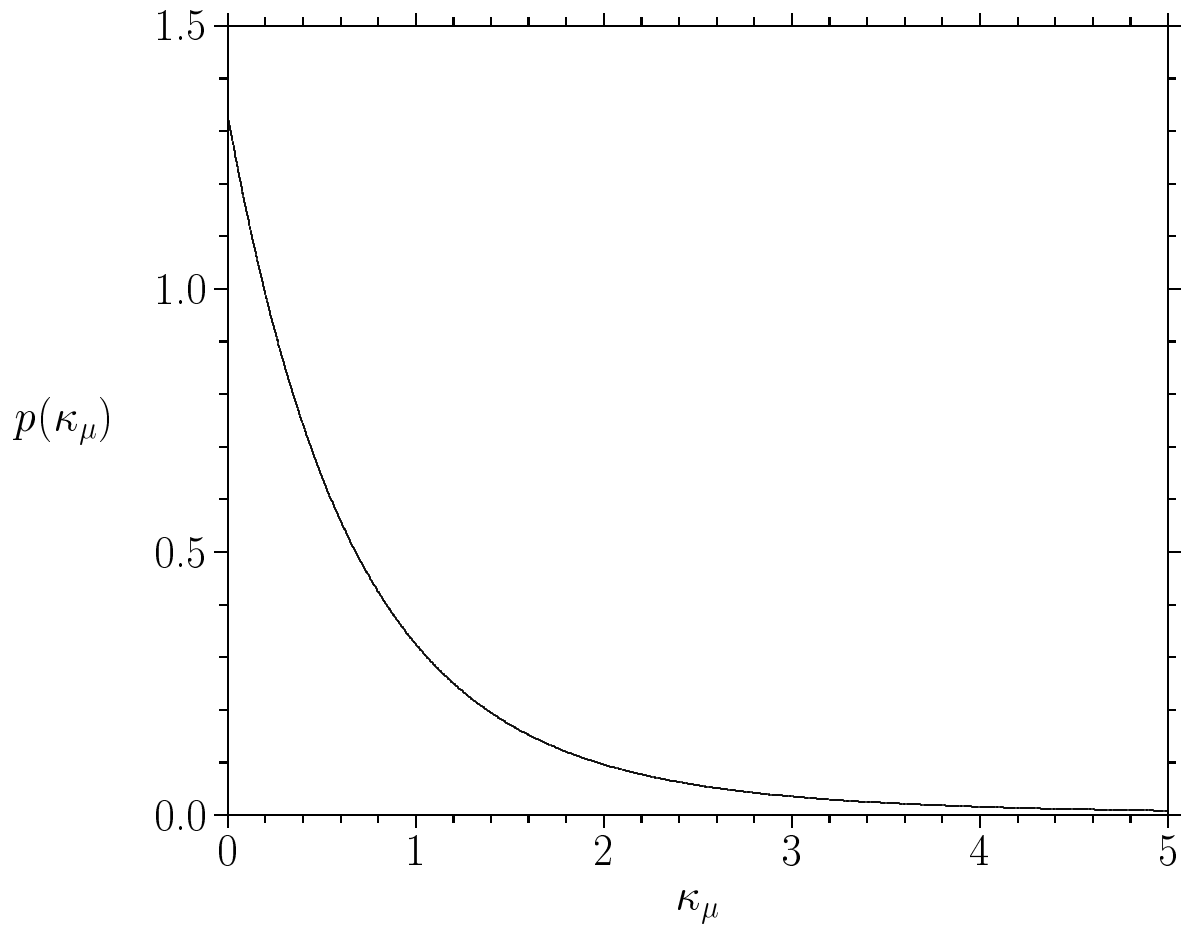
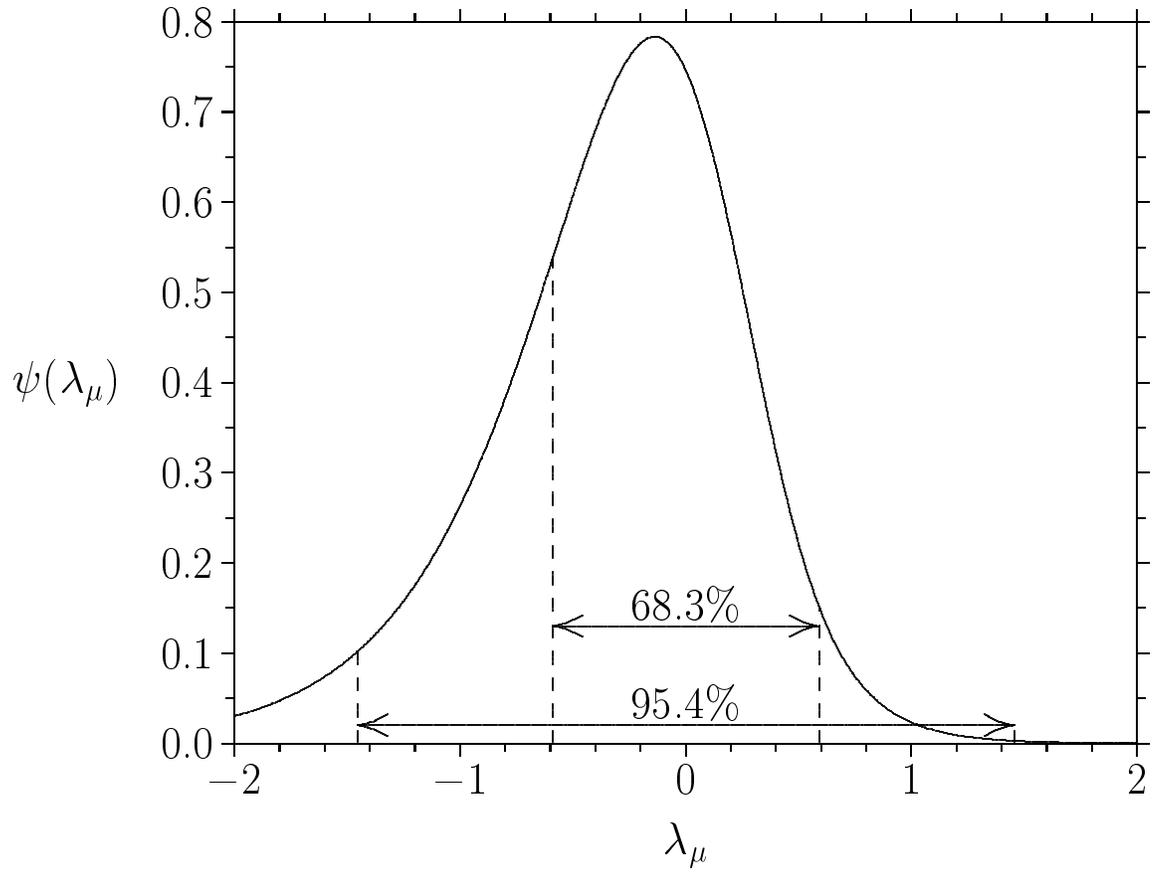
\begin{figure*}
\vspace{7cm}
\caption{The probability density $p(\kappa_{\mu})$
for $\kappa_{\mu} 
= \mu /\!<\!\!\mu\!\!>$ (left) and
the probability density $\psi(\lambda_\mu)$ 
with symmetric 68.3~\% and 95.4~\% intervals around $<\!\!\mu\!\!>$ (right)}
\label{halo:pkmu}
\end{figure*}

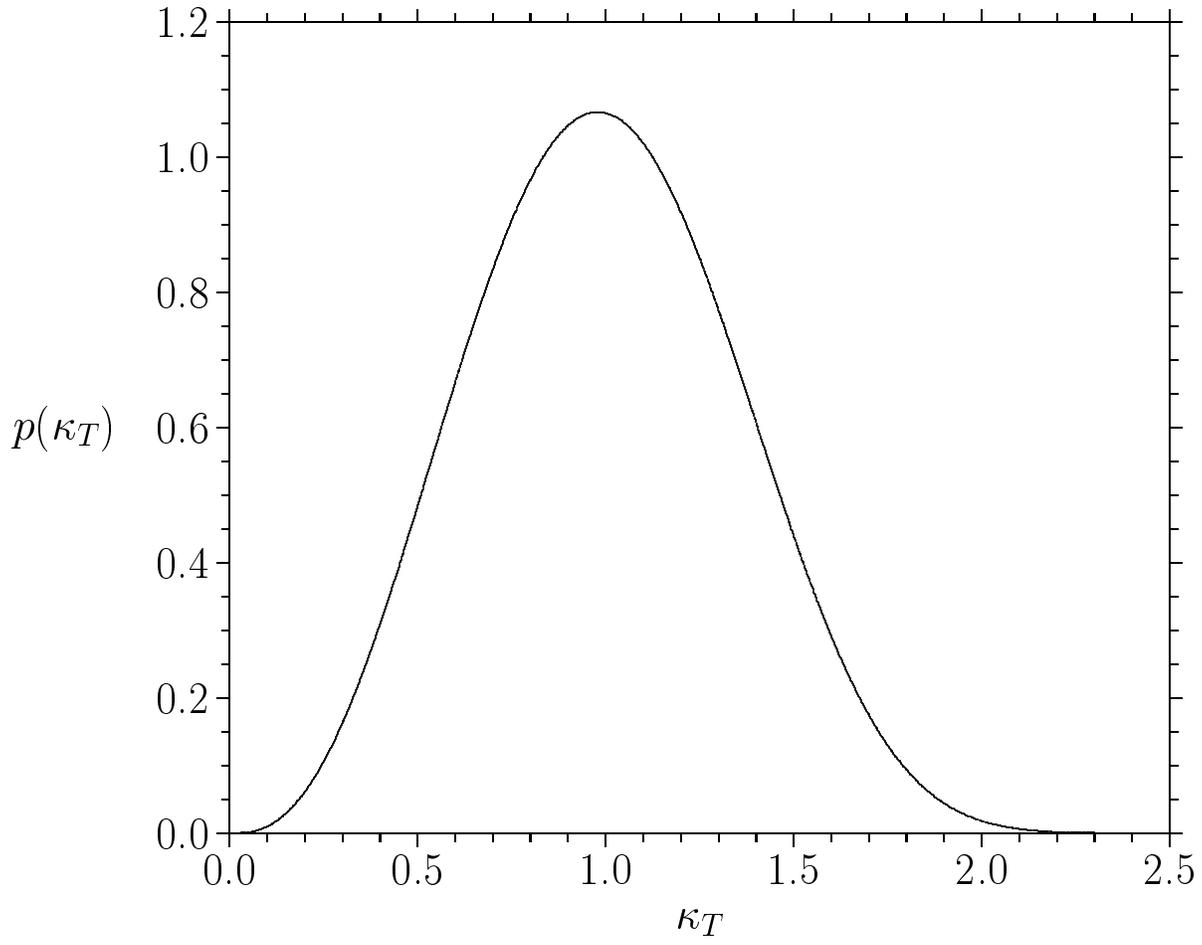
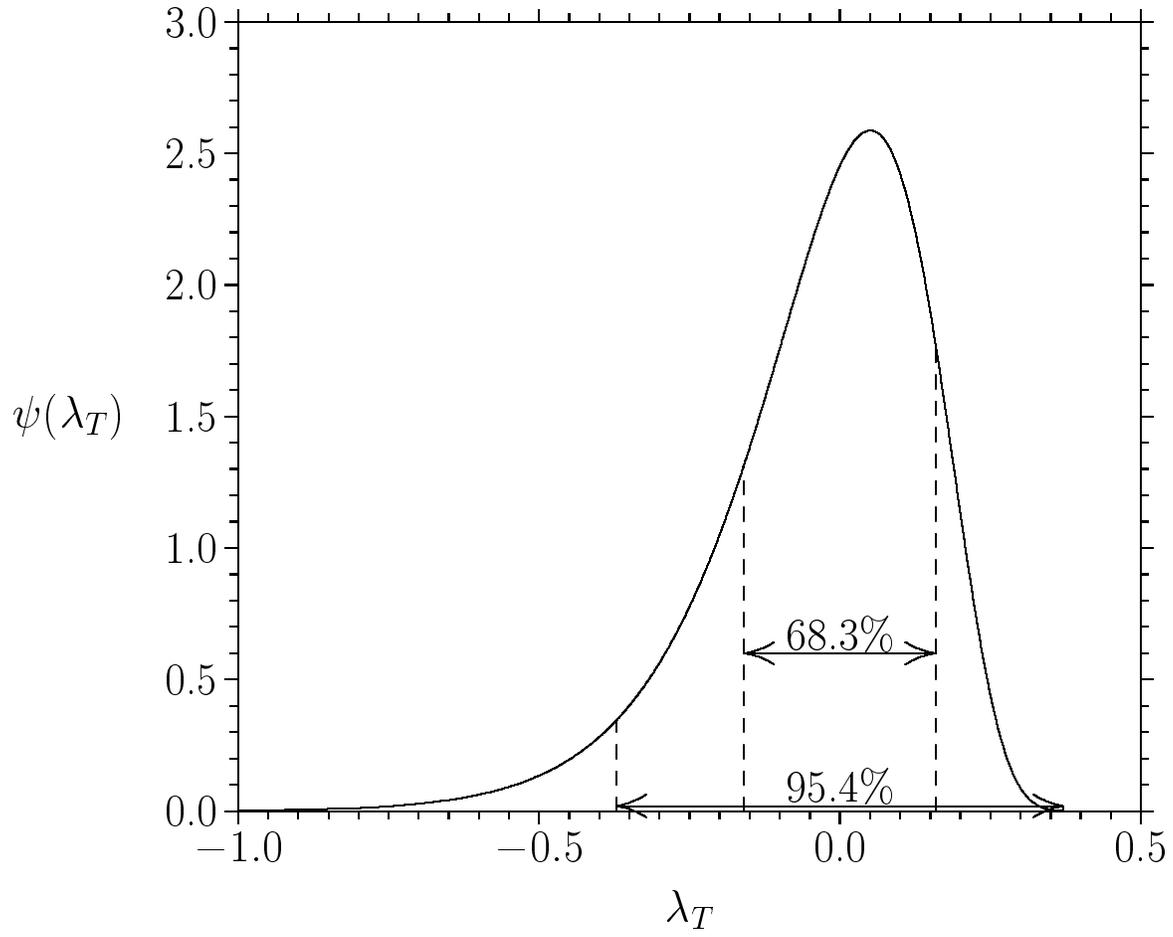
\begin{figure*}
\vspace{7cm}
\caption{The probability density $p(\kappa_T)$ for 
$\kappa_T = T /\!<\!\!T\!\!>$ (left) and
the probability density $\psi(\lambda_T)$ 
with symmetric 68.3~\% and 95.4~\% intervals around $<\!\!T\!\!>$ (right)}
\label{halo:pkt}
\end{figure*}

For the other quantities estimated the probability densities are given by
Eqs.~(\ref{ppsimasseq1}) and~(\ref{ppsimasseq2}).
The probability densities $p(\kappa_\mathrm{G})$ and $\psi(\lambda_\mathrm{G})$ for 
$r_\mathrm{E}$, $\mu$, $T$ are shown in Figs.~\ref{halo:pkre} 
to~\ref{halo:pkt}. Note that $r_\mathrm{E}$ follows the velocity 
distribution, because $k=0$. In the diagrams for $\psi(\lambda_G)$,
symmetric intervals around $<\!\!G\!\!>$ are shown which give a probability
of 68.3~\% and 95.4~\% respectively. The bounds of these intervals are
also shown in Table~\ref{halo:lambounds}.

\begin{table*}
\caption{The bounds of symmetric intervals on a logarithmic scale
around 
$<\!\!G\!\!>$ which correspond to probabilities of 68.3~\% and
95.4~\%}
\begin{flushleft}
\begin{tabular}{ccccccc}
\hline\noalign{\smallskip}
$G$ & $\Delta\lambda_{68.3}$ & $\Delta\lambda_{95.4}$ & 
$10^{-\Delta\lambda_{68.3}}$ & $10^{\Delta\lambda_{68.3}}$ & 
$10^{-\Delta\lambda_{95.4}}$ & $10^{\Delta\lambda_{95.4}}$ \\
\noalign{\smallskip}\hline\noalign{\smallskip}
$v_{\perp}$, $r_\mathrm{E}$ & 0.2428 & 0.6111 & 0.572 & 1.75 & 0.244 & 4.09 \\
$\mu$ & 0.5900 & 1.454 & 0.257 & 3.89 & 0.0351 & 28.5 \\ 
$T$ & 0.1588 & 0.3719 & 0.694 & 1.44 & 0.425 & 2.35 \\ \hline\noalign{\smallskip}
\end{tabular}
\end{flushleft}
\label{halo:lambounds}
\end{table*}
The bounds are much larger for $r_\mathrm{E}$ and again much larger for 
$\mu$ than for $T$, which is due to the wide distribution of
the velocity and $\mu \propto \zeta^2$, $r_\mathrm{E} \propto \zeta$,
while $T \propto \sqrt{\zeta}$.
The smallest and the largest value in the 95.4~\%-interval differ
by a factor of about 800 for $\mu$, 16 for $r_\mathrm{E}$ and 5 for
$T$.

\section{Application to observed events}
In this section I show the application of the method described here
to the observed events towards the
LMC. The first events have been claimed by EROS (Aubourg et al.\ \cite{aubourg}), namely
EROS\#1 and \#2, and
MACHO (Alcock et al.\ \cite{alcock1}), namely MACHO LMC\#1, in 1993. 
The fit with a point-mass lens and point source for
MACHO LMC\#1 showed a discrepancy near the peak which
has been solved with models involving a binary lens 
by Dominik \& Hirshfeld (\cite{dohi1}, \cite{dohi2}). 
The MACHO collaboration had found two other candidates, MACHO LMC\#2 and \#3,
(Alcock et al.\ \cite{alcock2}) which have been meanwhile dismissed. In addition,
they have claimed the existence of 7 additional events, MACHO LMC\#4 to \#10,
(Pratt et al.\ \cite{pratt}), where MACHO LMC\#9 is due to a binary lens 
(Bennett et al.\ \cite{bennett1}). In the MACHO data taken from 1996 to March 1997,
5 additional LMC candidates showed up (Stubbs et al.\ \cite{stubbs}).

It has been shown that the EROS\#2 event involves
a periodic variable star (Ansari et al.\ \cite{ansari}) and that
EROS\#1 involves an emission line Be type star (Beaulieu et al.\ \cite{beaulieu}), 
so that 
both EROS\#1 and EROS\#2
involve a rare type of stars which makes these events
suspect as microlensing candidates (e.g. Paczy{\'n}ski \cite{paczynski2}).
In addition, the MACHO LMC\#10 event 
is likely to be a binary star (Pratt et al.\ \cite{pratt}; Alcock et al.\ \cite{alcock4}).

By assuming that the lens is in the galactic halo and using the halo model
of the last section, expectation values for the desired quantities can be
obtained by inserting the fit parameters into Eqs.~(\ref{haloestfirst}) 
to~(\ref{haloestlast}).
Table~\ref{haloestsingle} shows the expectation values
for the Einstein radius and the mass for the events EROS\#1
and EROS\#2, MACHO LMC\#4\ldots{}\#8 and \#10, whereas
the results for the binary lens events MACHO LMC\#1 and MACHO LMC\#9
are shown in Table~\ref{estvalsbin}.
For MACHO LMC\#1 six different binary lens models are shown 
(Dominik \& Hirshfeld \cite{dohi2}; Dominik \cite{dothesis}). Note that the lens for
MACHO LMC\#9 probably resides in the LMC (Bennett et al.\ \cite{bennett1}).

\begin{table*}
\caption{Expectation values for the point-source-point-mass-lens
events towards the LMC. M \# denotes MACHO events, while E \# denotes
EROS events.}
\begin{flushleft}
\begin{tabular}{lccccccccc}
\hline\noalign{\smallskip}
\rule[-1ex]{0ex}{3.5ex} & M \#4 & M \#5 & M \#6 & M \#7 & M \#8 & M \#10 & E \#1 & E \#2 \\ 
\hline\noalign{\smallskip}
\rule[-1ex]{0ex}{3.5ex}$t_\mathrm{E}/\mathrm{d}$ & 
23 & 41 & 44 & 58 & 31 & 21 & 27 & 30  \\
\rule[-1ex]{0ex}{3.5ex}$<\!\!r_\mathrm{E}\!\!>/\mathrm{AU}$ & 
2.5 & 4.4 & 4.7 & 6.2 & 3.3 & 2.2 & 2.9 & 3.2 \\ 
\rule[-1ex]{0ex}{3.5ex}$<\!\!\mu\!\!>$ &
0.14 & 0.46 & 0.52 & 0.91 & 0.26 & 0.12 & 0.20 & 0.24 \\ 
 \noalign{\smallskip}\hline
\end{tabular}
\end{flushleft}
\label{haloestsingle}
\end{table*}

\begin{table*}
\caption{Expectation values of the physical parameters and used
fit parameters for the 6 
binary lens models for MACHO LMC\#1, 
denoted by BL, BL1, BA, BA1, BA2, and BA3, and 
for MACHO LMC\#9.}
\begin{flushleft}
\begin{tabular}{lccccccc}
\hline\noalign{\smallskip}
\rule[-1ex]{0ex}{3.5ex} & BL & BL1 & BA & BA1 & BA2 & BA3 & LMC\#9 \\ 
\noalign{\smallskip}\hline\noalign{\smallskip}
\rule[-1ex]{0ex}{3.5ex}$t_\mathrm{E}/\mathrm{d}$ & 
16.27 & 17.53 & 685 & 155 & 35.7 & $2.62\cdot 10^{12}$ & 143.4\\ 
\rule[-1ex]{0ex}{3.5ex}$t_\mathrm{E}^{(2)}/\mathrm{d}$ &
--- & --- & 17.57 & 15.15 & 17.72 & 16.36 & --- \\ 
\rule[-1ex]{0ex}{3.5ex}$m_1$ &
0.463 & 0.557 & 0.99934 & 0.9904 & 0.75 & $1-4\cdot 10^{-23}$ & 0.620 \\ 
\rule[-1ex]{0ex}{3.5ex}$r$ &
1.16 & 0.795 & $6.6\cdot 10^{-4}$ & $9.7\cdot 10^{-3}$ & 0.33 & 
$3.9\cdot 10^{-23}$ & 0.613 \\ 
\rule[-1ex]{0ex}{3.5ex}$\chi$ &
0.20 & 0.22 & 2.41 & 2.21 & 1.82 & 2.24 & 0.83 \\ 
\rule[-1ex]{0ex}{3.5ex}$\chi^{\,(1)}$ &
--- & --- & 2.41 & 2.22 & 2.10 & 2.24 & --- \\ 
\rule[-1ex]{0ex}{3.5ex}$<\!\!r_\mathrm{E}\!\!>/\mathrm{AU}$ & 
1.75 & 1.88 & 73.6 & 16.7 & 3.84 & $2.82\cdot 10^{11}$ & 15.3 \\
\rule[-1ex]{0ex}{3.5ex}$<\!\!r_\mathrm{E}^{(2)}\!\!>/\mathrm{AU}$ & 
--- & --- & 1.89 & 1.63 & 1.90 & 1.76 & --- \\ 
\rule[-1ex]{0ex}{3.5ex}$<\!\!2r_\mathrm{hd}\!\!>/\mathrm{AU}$ &
0.71 & 0.82 & 355 & 73.6 & 14.0 & $1.2\cdot 10^{12}$ & 2.54 \\ 
\rule[-1ex]{0ex}{3.5ex}$<\!\!\mu\!\!>$ &
0.072 & 0.083 & 127 & 6.5 & 0.34 & $1.9\cdot 10^{21}$ & 5.6 \\ 
\rule[-1ex]{0ex}{3.5ex}$<\!\!\mu_1\!\!>$ &
0.033 & 0.046 & 127 & 6.4 & 0.26 & $1.9\cdot 10^{21}$ & 3.5 \\ 
\rule[-1ex]{0ex}{3.5ex}$<\!\!\mu_2\!\!>$  &
0.038 & 0.037 & 0.084 & 0.062 & 0.086 & 0.073 & 2.1 \\ 
\rule[-1ex]{0ex}{3.5ex}$<\!\!T_\mathrm{min}\!\!>/\mathrm{a}$&
0.98 & 1.1 & 257 & 107 & 39 & $1.43\cdot 10^{7}$ & 24 \\ 
\noalign{\smallskip}\hline
\end{tabular}
\end{flushleft}
\label{estvalsbin}
\end{table*}

\begin{table}[htbp]
\caption{MACHO LMC\#1: Estimates of physical parameters of the binary lens models using
a mass ratio $r$ at the upper $2$-$\sigma$-bound 
of $r$ and corresponding values
of $t_\mathrm{E}^{\,(2)}$ and $\chi^{\,(1)}$.}
\begin{flushleft}
\begin{tabular}{lcccc}
\hline\noalign{\smallskip}
\rule[-1ex]{0ex}{3.5ex} & BA & BA1 & BA2 & BA3 \\ 
\hline\noalign{\smallskip}
\rule[-1ex]{0ex}{3.5ex}$t_\mathrm{E}/\mathrm{d}$ & 
57.67 & 86.93 & 29.94 & 228.6  \\
\rule[-1ex]{0ex}{3.5ex}$t_\mathrm{E}^{(2)}/\mathrm{d}$ &
17.63 & 15.20 & 17.61 & 16.49 \\ 
\rule[-1ex]{0ex}{3.5ex}$r$ & 
0.1031 & 0.0315 & 0.529 & 0.00523 \\ 
\rule[-1ex]{0ex}{3.5ex}$\chi$ & 
2.291 & 2.229 & 1.735 & 2.222 \\ 
\rule[-1ex]{0ex}{3.5ex}$\chi^{\,(1)}$ &
2.406 & 2.264 & 2.146 & 2.228 \\ 
\rule[-1ex]{0ex}{3.5ex}$<\!\!r_\mathrm{E}\!\!>/\mathrm{AU}$ & 
6.20 & 9.34 & 3.22 & 24.6 \\ 
\rule[-1ex]{0ex}{3.5ex}$<\!\!r_\mathrm{E}^{(2)}\!\!>/\mathrm{AU}$ & 
1.90 & 1.63 & 1.89 & 1.77 \\ 
\rule[-1ex]{0ex}{3.5ex}$<\!\!2r_\mathrm{hd}\!\!>/\mathrm{AU}$ &
28.4 & 41.7 & 11.2 & 109 \\ 
\rule[-1ex]{0ex}{3.5ex}$<\!\!\mu\!\!>$ &
0.90 & 2.0 & 0.24 & 14 \\ 
\rule[-1ex]{0ex}{3.5ex}$<\!\!\mu_1\!\!>$ &
0.82 & 2.0 & 0.16 & 14 \\  
\rule[-1ex]{0ex}{3.5ex}$<\!\!\mu_2\!\!>$  &
0.084 & 0.063 & 0.084 & 0.074 \\ 
\rule[-1ex]{0ex}{3.5ex}$<\!\!T_\mathrm{min}\!\!>/\mathrm{a}$&
69 & 82 & 33 & 132 \\ \noalign{\smallskip}\hline
\end{tabular}
\end{flushleft}
\label{estvalsbin2}
\end{table}

The parametrization used is that of Dominik \& Hirshfeld (\cite{dohi2}):
$r_\mathrm{hd}$ denotes the projected half-distance between the lens objects 
in the lens plane, where $r_\mathrm{hd} = 2\chi r_\mathrm{E}$, $m_1$ denotes
the mass fraction in lens object 1. $r_\mathrm{E}^{\,(2)}$ denotes the Einstein
radius corresponding to the mass of object 2 and $t_\mathrm{E}^{\,(2)}$ the
characteristic time corresponding to $r_\mathrm{E}^{\,(2)}$:
\begin{eqnarray}
r_\mathrm{E}^{\,(2)} = r_\mathrm{E}\,\sqrt{1-m_1}\,, \\
t_\mathrm{E}^{\,(2)} = t_\mathrm{E}\,\sqrt{1-m_1}\,.
\end{eqnarray}
Similarly $\chi^{\,(1)}$ denotes the projected half-separation in units
of Einstein radii corresponding to the mass of object 1, so that
\begin{equation}
\chi^{\,(1)} = \frac{1}{\sqrt{m_1}}\,\chi\,.
\end{equation}
$\mu_1$ and $\mu_2$ denote the mass in units of $M_{\sun}$ for objects 1 and
2 and the mass ratio $r$ is given by 
\begin{equation}
r = \frac{1-m_1}{m_1}\,.
\end{equation}
Since the true semimajor axis $a = \rho\, r_\mathrm{E}$ is not yielded by the
fit, $T$ is estimated using $\rho = \chi$, which corresponds to a minimal
value $T_\mathrm{min}$, because $\rho \geq \chi$ for any gravitationally bound
system and $T \propto \rho^{3/2}$.

The distribution of the physical quantities as well as symmetric intervals around
the expectation value with probabilities of 68.3~\% and 95.4~\% are
shown in the previous section.

The model BA3 has previously been omitted (Dominik \& Hirshfeld \cite{dohi2}) 
due to the fact that the mass
ratio between the lens objects is very extreme ($4\cdot 10^{-23}$), which 
corresponds to unphysical values (see Table~\ref{estvalsbin}). However, the
uncertainty of the mass ratio is very large for extreme mass ratios 
(Dominik \& Hirshfeld \cite{dohi2}) due to the fact that the lens behaves nearly
like a Chang-Refsdal lens. The same degenaracy has recently been 
rediscovered
in the context of lensing by a star with a planet by Gaudi \& Gould (\cite{GG}).
Table~\ref{estvalsbin2} shows the estimates for the 4 wide binary lens
models (BA, BA1, BA2, BA3) where fit parameters at the upper $2$-$\sigma$-bounds
of the mass ratio $r$ have been used. It can be seen that the expectation
values for the separation and the mass change dramatically for the fits
with small values of $r$, leaving much room for speculations about the
nature of the lens object.

Recently, Rhie \& Bennett (\cite{rhieben}) have speculated about a planetary companion
in the MACHO LMC\#1 event. However, as shown in this and the last section, 
there are two fundamental uncertainties (beyond the fact that there are only a few data 
points near the peak, which could have been solved by a denser sampling),
namely the uncertainty in the mass ratio and the uncertainty due to the 
unknown lens position and its velocity. The planetary model of 
Rhie \& Bennett (\cite{rhieben}) corresponds to my model BA1, where the expectation
value of the mass of the low-mass object is about $0.06~M_{\sun}$. 
Taking into account an uncertainty of a factor of 30, a value of 2~Jupiter
masses is reached just at the $2$-$\sigma$-bound. As shown in 
Table~\ref{estvalsbin}, the expectation value for the mass of the low-mass
object is about the same for all of my wide binary lens fits, so that there
are the same prospects for a planet as the low-mass object for all of these
models. However, the high-mass object will be different.

The estimates of physical quantities like the mass of the lens object(s),
their separation and the rotation periods along with the uncertainties
involved are needed to reveal the physical nature of
the lens for each observed event. One can check whether the mass range
is consistent with the assumption of a dark object, and of which nature
the lens should be in this case (brown dwarf, white dwarf, neutron star,
\ldots). The determination of the mass range is crucial for claiming the
existence of a planetary companion.

One can also check whether it is consistent to use
a static binary lens model rather than one with rotating binary lens
(Dominik \cite{do97rob}).
Moreover, by comparing estimates for different lens populations
(e.g. the galactic halo and the LMC halo) one may obtain indications 
to which population the lens belongs.

In contrast to estimates on the mass spectrum, there is no direct influence
from the ensemble of observed events on the estimates for a specific event.
However, there is an indirect influence, since the ensemble of events gives
information about the mass spectrum, which in turn can be used to get
a more accurate estimate for each observed event. However, a lot of events
($\sim 100$) are needed to get accurate information on the mass spectrum 
(Mao \& Paczy{\'n}ski \cite{MP2}). To be used for the estimates for a specific event,
some of the higher 
mass moments should have been determined. Note that there will
remain large uncertainties in the mass of a specific event even if the 
mass spectrum is known (unless it contains a sharp peak), since due to the
broad distribution of the velocity, the range of timescales $t_\mathrm{E}$ for
a certain mass is broad, so that the mass range for a specific event in turn 
may also be large.

While a white dwarf scenario is preferred by the LMC observations
(e.g. Pratt et al.\ \cite{pratt}),
a brown-dwarf scenario is not ruled out 
(Spiro \cite{spiro}) if one considers a flat halo. 
Though there are some restrictions on the average lens mass from the observed
events, which will improve with more data, it will remain highly uncertain
into which mass regime a certain event falls.

Since the ongoing microlensing observations constrain the galactic models
and therefore give
rise to more accurate determinations of the structure of the lens populations,
the knowledge on specific events will also be improved by this.

\begin{acknowledgements}
I would like to thank the MACHO collaboration for making available the
data for the MACHO LMC\#1 event, A.~C.\ Hirshfeld for a critical
reading of the manuscript, and P.~Jetzer for some discussion about the subject.

\end{acknowledgements}

\clearpage

\begin{figure*}
{\LARGE Figure 1 (left)}\\[5mm]

\epsfig{file=h0334.f1a}
\end{figure*}

\begin{figure*}
{\LARGE Figure 1 (right)}\\[5mm]

\epsfig{file=h0334.f1b}
\end{figure*}

\begin{figure*}
{\LARGE Figure 2 (left)}\\[5mm]

\epsfig{file=h0334.f2a}
\end{figure*}

\begin{figure*}
{\LARGE Figure 2 (right)}\\[5mm]

\epsfig{file=h0334.f2b}
\end{figure*}

\begin{figure*}
{\LARGE Figure 3 (left)}\\[5mm]

\epsfig{file=h0334.f3a}
\end{figure*}

\begin{figure*}
{\LARGE Figure 3 (right)}\\[5mm]

\epsfig{file=h0334.f3b}
\end{figure*}

\end{document}